\documentclass[aps,prl,reprint,superscriptaddress]{revtex4-1}
\usepackage{graphicx}
\usepackage{dcolumn}
\usepackage{bm}
\usepackage{amssymb}
\usepackage{amsmath}
\usepackage{microtype} 
\usepackage{amsfonts}
\usepackage{epsf}
\usepackage{epsfig}
\usepackage{color}
\usepackage{epstopdf}
\usepackage{xcolor}

\newcommand{\Fref}[1] {Fig.~\ref{#1}}

\newcommand{\ikf}{Institut f\"ur Kernphysik, Goethe-Universit\"at, Max-von-Laue-Strasse 1, 60438 Frankfurt, Germany}

\begin{document}
\title{Observation of Photoion Backward Emission in Photoionization of He and N$_2$}

\author{Sven~Grundmann}
\email{grundmann@atom.uni-frankfurt.de}
\address{\ikf}
\author{Max~Kircher} \address{\ikf}
\author{Isabel~V\'ela-Perez} \address{\ikf}
\author{Giammarco~Nalin} \address{\ikf}
\author{Daniel~Trabert} \address{\ikf}
\author{Nils~Anders} \address{\ikf}
\author{Niklas~Melzer} \address{\ikf}
\author{Jonas~Rist} \address{\ikf}
\author{Andreas~Pier} \address{\ikf}
\author{Nico~Strenger} \address{\ikf}
\author{Juliane~Siebert} \address{\ikf}
\author{Philipp~V.~Demekhin}
\address{
	\mbox{Institut f\"ur Physik und CINSaT, Universit\"at Kassel, Heinrich-Plett-Str. 40, 34132 Kassel, Germany}
}
\author{Lothar~Ph.~H.~Schmidt} \address{\ikf}
\author{Florian~Trinter} 
\affiliation{FS-PETRA-S, Deutsches Elektronen-Synchrotron (DESY), Notkestrasse 85, 22607 Hamburg, Germany}
\affiliation{Molecular Physics, Fritz-Haber-Institut der Max-Planck-Gesellschaft, Faradayweg 4-6, 14195 Berlin, Germany}
\author{Markus~S.~Sch\"offler} \address{\ikf}
\author{Till~Jahnke} \address{\ikf}
\author{Reinhard~D\"orner} \address{\ikf}

\date{\today}

\begin{abstract}
We experimentally investigate the effects of the linear photon momentum on the momentum distributions of photoions and photoelectrons generated in one-photon ionization in an energy range of 300~eV $\leq~E_\gamma~\leq$ 40~keV.
Our results show that for each ionization event the photon momentum is imparted onto the photoion, which is essentially the system's center of mass.
Nevertheless, the mean value of the ion momentum distribution along the light propagation direction is backward-directed by $-3/5$ times the photon momentum.
These results experimentally confirm a 90 year old prediction. 
\end{abstract}

\maketitle
The electric field vector of an electromagnetic wave is oriented perpendicular to the light propagation axis.
As this field drives photoionization, one might expect its direction to be the symmetry axis for angular distributions of photoelectrons and photoions.
At high photon energies $E_\gamma$ and corresponding high photon momenta $k_\gamma$, however, this symmetry is violated and the momentum distributions of reaction fragments are asymmetric with respect to the light propagation direction.
The observed forward/backward asymmetry of the emitted electrons has puzzled research for the last century (see \cite{Jenkin1981,Hemmers2004} for reviews).
In earliest photoionization studies, performed in 1927, Auger and Perrin left unanswered why their observed forward shift of photoelectrons was ``more than 50\% higher than the momentum of the photon $k_\gamma$'' \cite{Auger1927a}. 
Later, with the application of wave mechanics on photoionization, calculations reproduced the effect qualitatively and showed that it results from an interference between electric dipole and electric quadrupole transitions (e.g. \cite{Sommerfeld1930}), which both alone are forward/backward symmetric.
Already Sommerfeld and Schur realized \cite{Sommerfeld1930} that a mean forward momentum of electrons greater than the photon momentum ($\left\langle k^{e}_x \right\rangle > k_\gamma$) entails that the mean photoion momentum must be backward-shifted to account for momentum conservation.
This counter-intuitive prediction of backward-directed ions created by interaction with light that exerts a forward-directed radiation pressure stands experimentally untested till today.
In the present work we supply that missing evidence and demonstrate that the backward momentum of the ion scales with $-3/5 \cdot k_\gamma$ for a wide energy range of 300~eV -- 40~keV ($k_\gamma =$ 0.1 -- 12~a.u.). 

In a broader context, the so-called nondipole effects, which result from nonzero photon momentum, also have a significant impact on one-photon multiple ionization \cite{Briggs2000,Amusia1975,Schoffler2013,Grundmann2018,Chen2020}.
There, higher multipole components of the light-matter interaction do not only change the angular distribution of photoelectrons, but also open additional ionization pathways which are dipole-forbidden \cite{Maulbetsch1995}.
Recently, nondipole effects have also been studied for absorption of more than one photon \cite{Chelkowski2017} up to the extreme regime of strong-field tunnel ionization \cite{Smeenk2011,Chelkowski2014,Ludwig2014,Hartung2019}.
In the strong-field regime, the mechanism responsible for the radiation pressure changes.
Here, the action of the magnetic component of the field drives the electron forward, whereas the retardation of the electric field causes the symmetry breaking in the one-photon perturbative regime.

The experiments on one-photon ionization reported here have been performed at beamline P04 at PETRA~III (DESY, Hamburg, Germany \cite{Viefhaus2013}) using circularly polarized light (low-energy experiment, $E_\gamma=$ 300 -- 1775~eV) and beamline ID31 of the European Synchrotron Radiation Facility (ESRF, Grenoble, France) using linearly polarized light (high-energy experiment, $E_\gamma=$ 12 -- 40~keV).
We have used the ion arm of a COLTRIMS (Cold Target Recoil Ion Momentum Spectroscopy) reaction microscope \cite{Dorner2000, Ullrich2003, Jahnke2004a} to measure the charge state and the three-dimensional momentum vector of the photoions.
The photon beam was crossed at a right angle with a supersonic gas jet of He (low-energy experiment) or N$_2$ (high-energy experiment).
The ions were guided by an electric field to a time- and position-sensitive detector with delay-line position readout \cite{Jagutzki2002,Jagutzki2002a}.
From the ions' times-of-flight and positions-of-impact the initial momenta after photoionization were retrieved.
For the case of N$_2$ we considered only $K$-shell ionization followed by Auger decay.
In this case, two singly charged ions are created which we detected in coincidence with the Auger electron.
From these three momentum vectors we calculated the momentum of the N$_2^{+}$ ion at the instance after photelectron emission.
These results on N$_2$ are gained from the same experimental runs as \cite{Kircher2019,Kircher2019a}, where further experimental details can be found.
Our spectrometer captured ions with $4\pi$ collection solid angle, which allowed to directly obtain the mean value of the momentum from the data.
To access the ion momenta on an absolute scale, it is essential to know precisely the location of ions with zero momentum on our detector.
For the high-energy data, this zero point is obtained from ions which are created by Compton scattering.
In this case, the photon momentum is transferred to the electron and the ion is thus left with a momentum distribution centered at the origin \cite{Spielberger1995}.
For the low-energy data we extrapolate the zero as we show in more detail in Fig.~\ref{fig2}.

Figure~1 summarizes the results of our study.
Shown in blue is the measured mean value of the ion momentum in direction of the light propagation $\left\langle k_x^{ion} \right\rangle $ as function of photon energy (top scale) or photon momentum (bottom scale).
The full circles (low photon energies) correspond to single ionization of He, the full squares (high photon energies) to $K$-shell ionization of $\text{N}_2$.
In the latter case, as outlined before, the measured mean value of the sum momentum of both ionic fragments is adjusted by the momentum of the Auger electron.
Negative values correspond to backward emission, i.e. against the photon propagation direction.
The corresponding mean value of the photoelectron momentum $\left\langle k_x^{e} \right\rangle $, which is obtained from the measured ion momentum by using momentum conservation ($\left\langle k_x^{e} \right\rangle = - \left\langle k_x^{ion} \right\rangle + k_\gamma$), is plotted in red.
The red and blue lines show the corresponding prediction from \cite{Chelkowski2014}:
\begin{eqnarray}
\left\langle k^{ion}_x \right\rangle = -\frac{3}{5} \frac{E_\gamma - I_p}{c} + \frac{I_P}{c} = -\frac{3}{5} k_\gamma \bigg\rvert_{I_p=0} 
\end{eqnarray}
and
\begin{eqnarray}
\left\langle k^{e}_x \right\rangle = \frac{8}{5} \frac{E_\gamma - I_p}{c} = \frac{8}{5} k_\gamma \bigg\rvert_{I_p=0}
\end{eqnarray}
\noindent
where $I_P$ is the ionization potential and $c$ is the speed of light.
Our finding thus yields direct experimental confirmation of the long standing prediction of backward-directed ion emission in photoionization \cite{Sommerfeld1930,Michaud1970,Seaton1995,Chelkowski2014}.

\begin{figure} [t]
\centering
\includegraphics[width=1.\columnwidth]{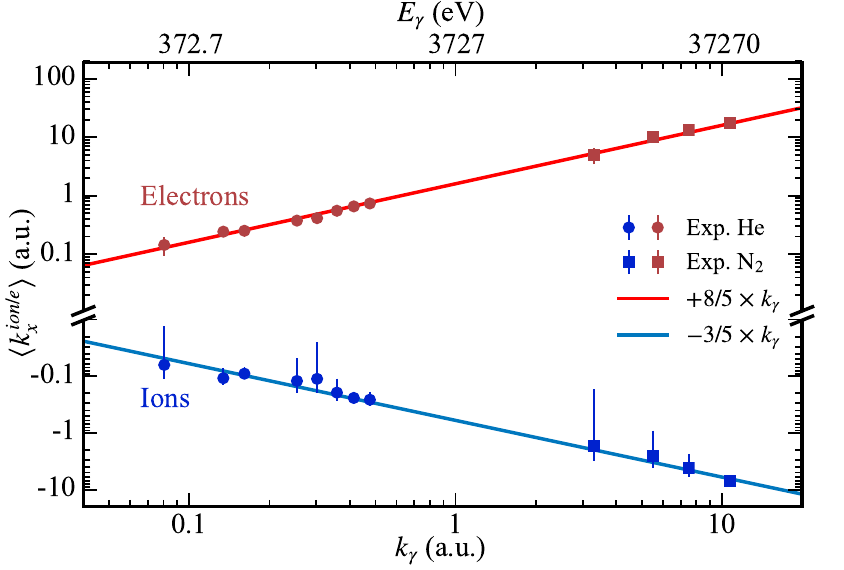}
\caption{Mean value of electron (red) and ion (blue) momenta along the light propagation axis after one-photon ionization. Horizontal axis: Photon momentum $k_\gamma=E_\gamma/c$ (bottom scale) and photon energy (top scale). Circles: He photoionization (circularly polarized light); squares: N$_2$ $K$-shell photoionization followed by Auger decay, where the mean value corresponds to the sum of both N$^+$ fragments adjusted by the Auger electron (linearly polarized light). The lines show Eqs.~(1) and (2), respectively. Note that in order to compare the He and the N$_2$ datasets, $I_P$ was approximated as zero in this figure.
The error bars are asymmetric only because the y-axis is logarithmic.
} 
\label{fig1}
\end{figure}

A fully differential view of the momentum distributions, which underlie the mean value, clearly visualizes the physics origin of the observed ion backward emission.
Figure 2 shows the measured photoion momentum distributions for photoionization of He by 300, 600, 1125, and 1775~eV circularly polarized photons in cylindrical coordinates.
The horizontal axis is the momentum component parallel to $k_\gamma$, the vertical axis is the momentum perpendicular to the photon axis. By definition this momentum is positive, we have mirrored the distribution at the horizontal axis.
The datasets from the different photon energies are normalized to result in the same maximum intensity for better visibility.
\begin{figure} [t]
\centering
\includegraphics[width=1.\columnwidth]{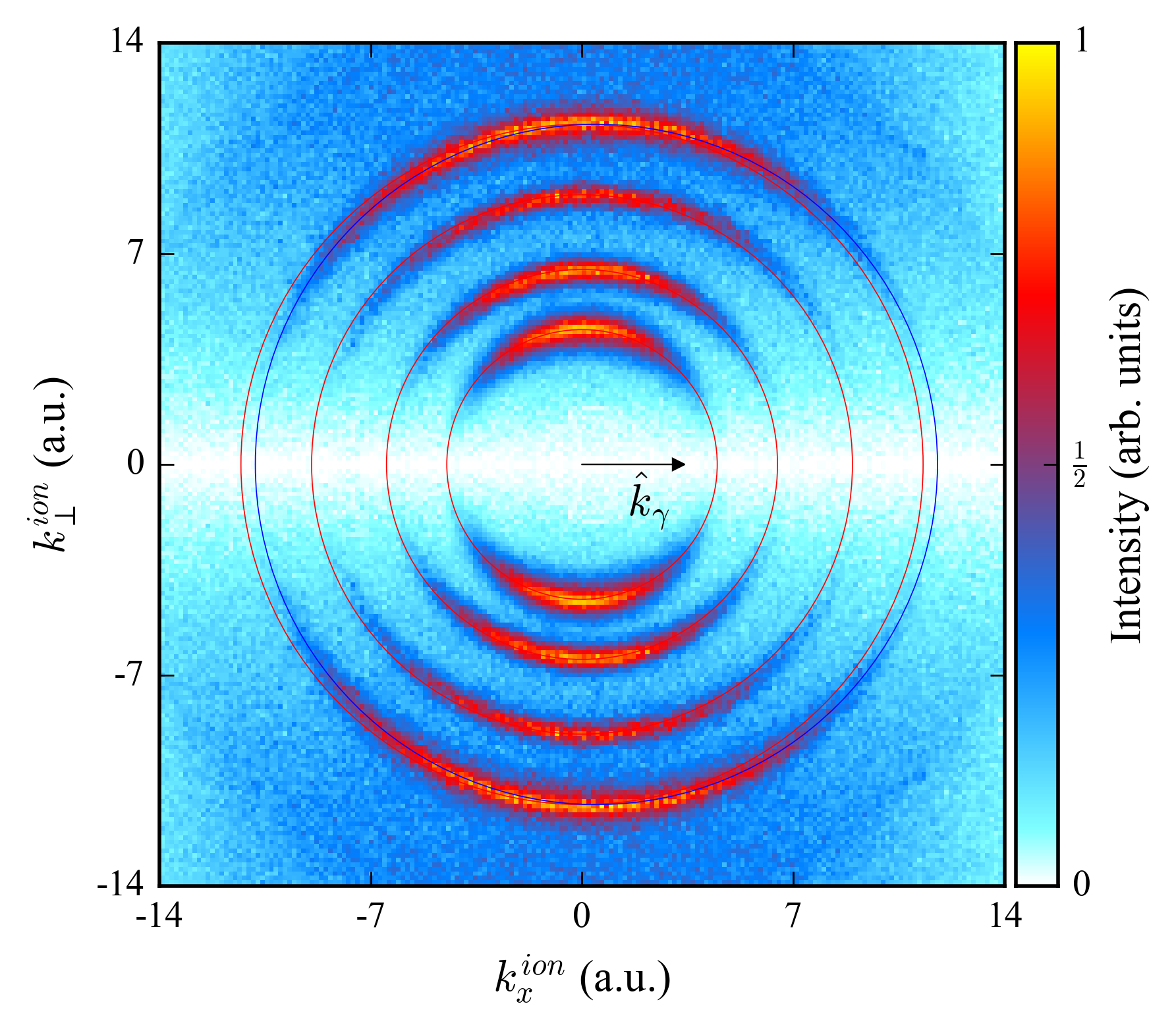}
\caption{Momentum distribution of He$^+$ ions from single ionization by circularly polarized photons with $E_\gamma=$ 300, 600, 1125, and 1775~eV (from inner to outer ring). Horizontal axis: momentum component parallel to light propagation axis. Vertical axis: momentum perpendicular to the light propagation. The lower half is a mirror image of the upper half of the figure. The red circles are centered at the origin with a radius of the electron momentum $k^e = \sqrt{2(E_\gamma - I_P)}$. The blue outer circle is forward-shifted by the corresponding photon momentum $k_\gamma=0.476$~a.u. The data from the different photon energies are arbitrarily normalized to match the color scale.}
\label{fig2}
\end{figure}
The red circles represent concentric rings centered on the origin in momentum space.
Their radii correspond to the respective photoelectron momenta $k^e = \sqrt{2(E_\gamma - I_P)}$.
The events do not accumulate on these rings, but are shifted forward in direction of photon propagation.
This is most clearly visible at the outermost ring corresponding to 1775~eV photon energy.
The shift of the experimental distributions is given by the respective photon momentum.
To guide the eye, the blue circle is shifted forward by the photon momentum of a 1775~eV photon (0.476~a.u.).
Hence, the measured ion momentum distributions directly show that the photon momentum is mostly absorbed by the ion, which is strictly a consequence of momentum conservation.
In each individual ionization event, the photon momentum is transferred to the center of mass of the system, which almost coincides the ion.
Accordingly, the corresponding momentum distribution of the electron shows a circle of the same radius, but not forward shifted.
A full derivation of the kinematics including higher order corrections can be found in \cite{Dorner2000}.

Besides the forward shift of the ring in the ion momentum space, also the distribution of counts on that ring changes with photon energy. This distribution is tilted more to the backward hemisphere upon increase of $E_\gamma$.
By momentum conservation, the ion's final momentum is the photon momentum (causing the forward shift of the sphere) minus the photoelectron momentum.
Thus, the distribution of the ions on the shifted sphere in momentum space is a direct mirror image of the angular distribution of the photoelectrons in the laboratory frame. 
\begin{figure} [t]
\centering
\includegraphics[width=1.\columnwidth]{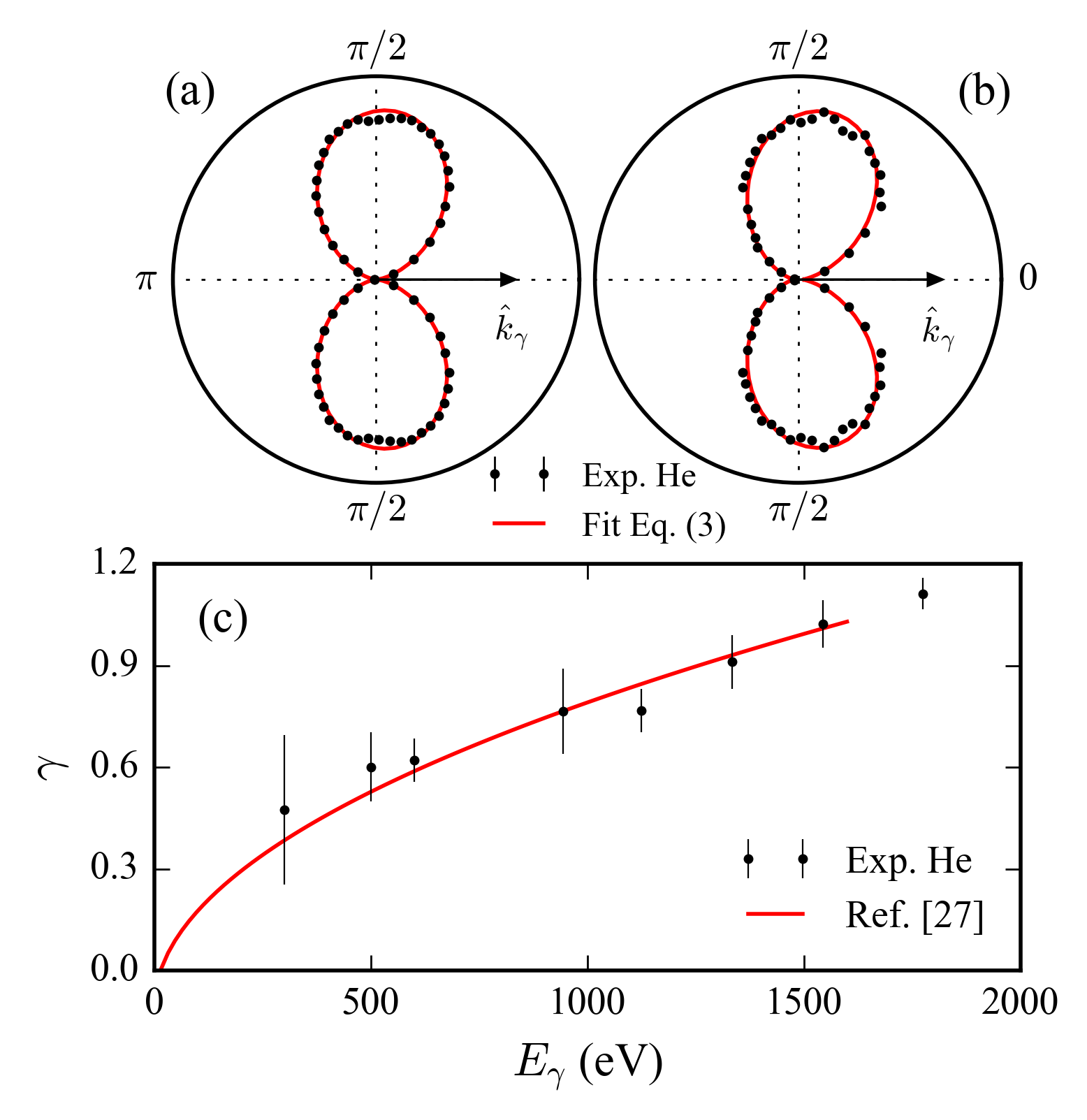}
\caption{Electron angular distributions after single ionization of He obtained from the measured ion momentum distributions shown in \Fref{fig2}, using $\boldsymbol k^e=-\boldsymbol k^{ion}+\boldsymbol k_{\gamma}$. (a) $E_\gamma$ = 300 eV, (b) $E_\gamma$ = 1775 eV. Red line: Fit using Eq.~(3). (c) Nondipole parameter $\gamma$ from fit of Eq.~(3) to the data. Red line: Calculations from Ref. \cite{Amusia2001}.}
\label{fig3}
\end{figure}
These electron angular distributions obtained from our measured ion momentum distributions are shown in \Fref{fig3}.
They have an approximate dipolar shape as the initial state is He(1$s$) and thus the leading angular momentum term in the final state is a dipole ($l=1$).
In addition, this dipolar shape is forward-tilted.
In order to characterize this effect by an appropriate differential cross section, the dipole approximation has to be extended by a leading first-order correction, i.e. the interference term between the electric dipole (E1) and quadrupole transitions (E2).
Accordingly, the angle-differential cross section for photoionization can be written as
\begin{equation}
\frac{d\sigma}{d\vartheta}\propto 1- \beta~ \frac{3 \cos^2\vartheta-1}{4}+\gamma~\frac{\sin^2\vartheta \cos\vartheta}{2}
\end{equation}
for circularly polarized (and unpolarized) light \cite{Cooper1990, Cooper1993}.
Here, $\beta$ is the dipole anisotropy parameter, $\gamma$ is the nondipole parameter characterizing the E1--E2 interference, and $\vartheta$ is the angle enclosed by photon and photoelectron momenta. 
Angle-differential cross sections (and thus the parameters $\beta$ and $\gamma$) are shaped by the coherent superposition of all possible angular momentum channels of the photoelectron.
For a He atom, the dependence of $\gamma$ on the energy is well reproduced by theory.
Our data are in good agreement with published data from Ref. \cite{Amusia2001}.

There are many misleading formulations on the photon momentum transfer in the literature.
Often, it is sloppily stated that the ``absorbed photon imparts its own momentum to the ejected electron'' \cite{Bethe1957} suggesting that this ``kick'' is responsible for the forward tilt of electron angular distributions as shown in Fig. \ref{fig3}.
For a more accurate formulation, it is instructive to recapitulate how the photon momentum transfer emerges from the interaction with the electromagnetic field.
For simplicity, we consider photoionization of the $1s$ electron of the hydrogen atom and follow the formalism of \cite{Felicisimo2005}.
Beyond the electric dipole approximation, an ionizing plane electromagnetic wave with the wavevector $\vert\boldsymbol k_\gamma\vert=k_\gamma=E_\gamma/c$ (photon momentum) imprints the local phase factor $e^{i\boldsymbol k_\gamma\cdot \boldsymbol{r}}$ in the total transition matrix element.
By introducing the coordinate $\boldsymbol R_H$ for the center of mass of the atom and the coordinate $\boldsymbol r^\prime$ for the $1s$ electron with respect to $\boldsymbol R_H$, the absolute coordinate of the $1s$ electron in the laboratory frame can be rewritten as $\boldsymbol{r}=\boldsymbol R_H+\boldsymbol r^\prime$. The respective phase can thus be factorized as follows:
\begin{equation}
\label{eq:phase}
e^{i\boldsymbol k_\gamma \cdot \boldsymbol{r}}=e^{i\boldsymbol k_\gamma \cdot \boldsymbol R_H}  \, e^{i\boldsymbol k_\gamma\cdot \boldsymbol r^\prime}.
\end{equation}
This phase, introduced by the field, modifies the full transition matrix element: The first factor from Eq.~(\ref{eq:phase}) enters the transition matrix element  $\langle \boldsymbol{\pi}\vert e^{i\boldsymbol k_\gamma \cdot \boldsymbol R_H} \vert \boldsymbol{\pi}_0 \rangle$ between the translational states of the atomic center of mass, which are described by the plane waves $(2\pi)^{-3/2}e^{i\boldsymbol{\pi}\cdot \boldsymbol R_H}$ with  momentum $\boldsymbol{\pi}$.
This amplitude yields the momentum conservation law $\boldsymbol{\pi}=\boldsymbol{\pi}_0+\boldsymbol k_\gamma$.
Thus, a photon absorption by the atom introduces to its center of mass the momentum  $\boldsymbol k_\gamma$.
The second phase factor $e^{i\boldsymbol k_\gamma \cdot \boldsymbol r^\prime}$ from Eq.~(\ref{eq:phase}) enters the electric dipole transition matrix element and is responsible for the multipole (retardation) corrections beyond the electric dipole approximation.

The photon momentum and nondipolar photoionization are of relevance in many areas such as astrophysics of stellar outer layers \cite{Seaton1996} or for acceleration of electrons by relativistic laser pulses \cite{DiPiazza2012}.
The simplest and most transparent process where the photon momentum manifests is one-photon excitation where the atom receives the photon momentum, providing, for example, a tool for laser cooling.
Above the ionization threshold in each ionization event, the ion receives the photon momentum and in addition the recoil of the photoelectron.
The additional orbital angular momentum transfer from the photon leads to an increasing forward tilt of the electron angular distribution.
This forward-directed mean momentum of the electron is balanced by a backward-directed momentum transfer to the ion.
Our results show that for $s$-initial states, the ion backward momentum scales with $-3/5 \cdot k_\gamma$, confirming a long standing prediction.
For more complex processes like double ionization \cite{Chen2020}, the photon momentum sharing is less straightforward in detail, the ion backward emission, however, prevails.
This is very different for multi-photon processes in strong laser pulses where electrons are forward-driven by the magnetic component of the light field \cite{Chelkowski2017,Smeenk2011,Chelkowski2014,Ludwig2014}.
Our present observation on ion momenta can be extended in the future to ionization in the multi-photon regime in order to test the prediction that high fields direct the ions forward \cite{Chelkowski2014}.

\begin{acknowledgments}
We acknowledge DESY (Hamburg, Germany), a member of the Helmholtz Association HGF, for the provision of experimental facilities.
Parts of this research were carried out at PETRA III and we would like to thank J\"orn Seltmann and Kai Bagschik for excellent support during the beam time.
We are grateful for the staff at ESRF for providing the high-energy photon beam and thank V.~Honkim\"aki, J. Drnec, H. Isern, and F. Russello from beam line ID31 at ESRF for excellent support during the beam time.
We acknowledge support by DFG and BMBF.
We acknowledge support of the theory-experiment collaboration by DFG via SFB 1319 (ELCH).

\end{acknowledgments}
%
%
%
\bibliography{library}
\end{document}